\documentclass[pra,twocolumn]{revtex4}
\usepackage[koi8-r]{inputenc}
\usepackage[english]{babel}
\usepackage{amsmath}
\usepackage{amssymb}
\usepackage{epsfig}

\begin{document}

\title{Three-dimensional numerical simulation of long-lived quantum vortex knots 
and links in a trapped Bose-Einstein condensate}
\author{V.~P. Ruban}
\email{ruban@itp.ac.ru}
\affiliation{L.D. Landau Institute for Theoretical Physics RAS,
142432 Chernogolovka, Moscow region, Russia} 
\date{\today}

\begin{abstract}
Dynamics of simplest vortex knots, unknots, and links of torus type inside an atomic
Bose-Einstein condensate in anisotropic harmonic trap at zero temperature has 
been numerically simulated using three-dimensional Gross-Pitaevskii equation. 
The lifetime for such quasi-stationary rotating vortex structures has been found 
quite long in wide parametric domains of the system. This result is in qualitative
agreement with a previous prediction based on a simplified one-dimensional model 
approximately describing dynamics of vortex filaments [V.P. Ruban, JETP {\bf 126}, 397 (2018)].
\end{abstract}
\maketitle

{\bf Introduction}.
The study of dynamics and statics of quantized vortex filaments has been one of the 
most important directions in the theory of cold Bose-Einstein-condensed atomic gases
(see review \cite{F2009} and citations therein).
It is usually distinguished between vortex rings
(those are closed vortices with simple topology), vortex filaments in the narrow sense
(when ends of vortices start and finish on the surface of condensate cloud), and
topologically non-trivial configurations as vortex knots and links. Behavior of the system
is strongly influenced by the fact that condensate is placed in an external trapping potential,
so the vortex motion takes place on a highly nonuniform spatial density background
$\rho({\bf r})$, and in a very confined space. Practically, the ratio of the system size
(condensate cloud size) to a typical vortex core width does not exceed a few tens. 
In this sense, the situation differs strongly from, say, quantum vortex turbulence in 
helium-II where the mentioned ratio can reach eight orders of magnitude while the superfluid 
density background is highly homogeneous. For moderate-size Bose-Einstein condensates,
long-lived configurations with a single or just a few vortices are of interest.
A lot of research has been carried out about rings and filaments in nonuniform condensates
(see, e.g., 
\cite{SF2000,FS2001,R2001,AR2001,GP2001,AR2002,RBD2002,AD2003,AD2004,D2005,
Kelvin_vaves,ring_istability,v-2015,BWTCFCK2015,R2016-1,R2016-2,R2017-2,R2017-3,
reconn-2017,top-2017,WBTCFCK2017,TWK2018}), while the knot dynamics was studied up to now
mainly for uniform density
(see \cite{RSB999,MABR2010,POB2012,KI2013,POB2014,LMB016,KKI2016,R2018-2,R2018-3},
and references therein). An experimental technique for producing knots and links 
in condensates has not yet been developed too. Only very recently in Ref.\cite{R2018-1},
based on hydrodynamic approximation (with potential perturbations neglected), simple
vortex knots in trapped axisymmetric condensates have been theoretically considered.
A shape of each such knot in the cylindrical coordinates is described by a complex-valued,
$2\pi P$-periodic on the azimuthal angle $\varphi$ function
$A(\varphi,t)=Z(\varphi,t)+iR(\varphi,t)$.
It was predicted that if the maximum of function $r\rho(z,r)$ is locally isotropic 
[the matrix of second partial derivatives is proportional to the unit $2\times 2$ matrix 
at the extremum point $(Z_*,R_*)$], and if the local induction parameter is large,
$\Lambda=\ln(R_*/\xi_*)\gg 1$ [where $\xi_*$ is the vortex core width corresponding 
to equilibrium density $\rho_*=\rho(Z_*,R_*)$], then sufficiently ``low-amplitude'' torus
vortex knots should be stable in such condensate. Applicability of the theory is 
limited by the conditions 
\begin{equation}
\xi_*\ll |A_n-A_m| \ll R_*, \quad |A_n-A_*|\ll R_*,
\end{equation}
where $A_n(\varphi,t)=A(\varphi+2\pi n,t)$, $n=0,1,\dots, P-1$. Let us choose the length  
scale and the origin so that $R_*=1$, $Z_*=0$. The stability prediction for simplest torus
knots follows from the approximate system of equations which determine the dynamics
of functions $W_n(\varphi,t)=\sqrt{\Lambda}(A_n-i)$ in the case when the mentioned matrix
of second derivatives is  $-$Diag$(\alpha-\epsilon,\alpha+\epsilon)$ 
(see details of derivation in Ref.\cite{R2018-1}):
\begin{equation}
\frac{i}{\Lambda}W_{n,t}=-\frac{1}{2}[W_{n,\varphi\varphi}+\alpha W_n-\epsilon W_n^*]
-\sum_{j\neq n}\frac{1}{W_n^*-W_j^*}.
\label{W_equations}
\end{equation}
The same system but with different boundary conditions at $\varphi=0$ and $2\pi$ is able to
describe the motion of several vortices with total number $P$ of turns around $z$ axis.
It is possible to check directly that in the case $\epsilon=0$ at $P\leq 6$, all the solutions
of the form
\begin{eqnarray}
A_n&=&i+C_0\exp(i\alpha\Lambda t/2)\nonumber\\
&+&B_0\exp\Big(\frac{2\pi i n}{P}+\frac{iQ}{P}\varphi-it\Omega (Q,P,B_0)\Big),
\label{torus}
\end{eqnarray} 
with constant $B_0$ and $C_0$, are stable. With $C_0=0$, such solutions correspond 
to stationary rotating torus knots --- if the integers $Q$ and $P$ are co-prime,
``unknots'' --- if $|Q|=1$, or links --- when $Q$ and $P$ have common multipliers. 

At non-zero $\epsilon$, instability bands appear in the parameter plane $(\alpha, B_0)$,
being narrow accordingly to smallness of $\epsilon$.

It is clear that a real picture cannot be so refined as described above, 
especially in the case of not too large $\Lambda$ and taking into account 
terms that were neglected in derivation of Eqs.(\ref{W_equations}).
Moreover, it should not be forgotten that vortices interact with potential degrees of freedom.
Therefore the best we can expect in a trapped condensate is a rather long but still 
finite lifetime of vortex knots and links in a quasi-stationary regime.

The possibility of existence of long-lived vortex structures of the indicated type in real
Bose-Einstein condensates seems as a hypothesis deserving serious attention.
The purpose of the present work is its numerical verification in the framework
of Gross-Pitaevskii equation as a more accurate three-dimensional (3D) model
for a rarefied Bose gas at zero temperature. To the best author's knowledge, 
contrary to spatially uniform systems, a direct 3D simulation of torus vortex knots 
and links in a trapped condensate was not performed previously.
 
{\bf Numerical method}.
Let for simplicity the trap potential be quadratic. The properly non-dimensionalized 
Gross-Pitaevskii equation for the condensate wave function has the form 
\begin{equation}
i\psi_t=-\frac{1}{2}\Delta\psi 
+\frac{1}{\xi^2}\Big[\frac{1}{2}(x^2+y^2+az^2)-\frac{3}{2}+|\psi|^2\Big]\psi.
\label{GP}
\end{equation}
Only two dimensionless parameters are present here, a trap anisotropy $a$ and a small
healing length (typical vortex core width) $\xi\ll 1$. We note that inside the condensate, 
an approximate equality takes place,
\begin{equation}
\rho(z,r)\approx 3/2-(r^2+az^2)/2.
\end{equation}
Ellipsoid $r^2+az^2=3$ is a conditional boundary of the condensate.
Calculation of second partial derivatives for $r\rho(z,r)$ at point $Z_*=0$, $R_*=1$ gives
$\alpha=(3+a)/2$, $\epsilon=(3-a)/2$, i. e. parameter $\epsilon$ is equal to zero when $a=3$.
That corresponds to an oblate cloud with the size ratio $1/\sqrt{3}$.

Equation (\ref{GP}) was simulated numerically with a uniform spatial resolution as 
$256^3$ points in a cubic domain  $(2\pi/1.6)^3$ centered at the origin, and with periodic
boundary conditions. The indicated cube size ($L\approx 4$) ensured a sufficient decay
of wave function near the cube boundary and simultaneously, the condensate having
a transverse  size about $2\sqrt{3}\approx 3.5$ occupied considerable portion of the 
computational domain. Thus, the spacing between neighboring grid points was about $1/60$.
Since parameter $\xi$ must be as long as several grid periods, and on the other hand, 
$\xi$ must be small enough for possibility of comparison of numerical results to the 
analytical theory, the value $1/\xi^2=200\cdot 2.56$ was adopted which corresponded to 
$\xi\approx 1/22.5$ and the local induction parameter $\Lambda\approx 3$.

For time-stepping, the widely applied for such systems Split-Step Fourier Method 
of the second order of approximation was used. The Hamiltonian functional of the system,
as well as its wave-action integral, were conserved up to 4 decimal places over the simulation 
time  $T_{\rm sim}=80/2.56 >30$.

Initial shape of the vortex was determined by expression (\ref{torus}) with $t=0$, $C_0=0$,
and with adding to $B_0$ small azimuthal perturbations: 
\begin{equation}
B_0\longrightarrow B_0+0.002\cdot\sum_{m=1}^5 \cos(m\varphi+\gamma_m),
\end{equation}
where $\gamma_{1,2,3,4,5}=3,5,8,13,21$. These perturbations break the symmetry and serve as
additional ``seeds'' for development of possible instabilities.

Besides that, preliminary simulations have shown that in order to ensure a quasi-stationary 
regime, it is better in the right-hand side of Eq.(\ref{torus}) to use instead of the first 
term $i$ a somewhat smaller quantity, approximately $0.9 i$. That corresponds to a smaller 
torus radius than the theory predicts. Presumably, the decreased radius is due to a shift
of the maximum position of function $\Lambda(z,r)r\rho(z,r)$ which describes the energy
of a coaxial vortex ring, and the multiplier $\Lambda$ cannot be considered as 
almost constant at moderate values $\Lambda\sim 3$, contrary to the derivation of
approximate equations (\ref{W_equations}) in Ref.\cite{R2018-1}.

In order to prepare an initial complex field $\psi_0=\psi(x,y,z, t=0)$ containing the vortex
but with potential excitations as weak as possible, a special procedure was designed.
In its first stage, $\psi_0$ was determined by a more-less appropriate analytical formula which
roughly corresponds to the condition $|\psi_0|^2\approx 3/2-(r^2+az^2)/2$, but exactly
corresponds to the presence of phase singularity, i.e. vortex knot of a prescribed shape.
The spatial phase distribution, naturally, was far from being optimal at that state.
In the second stage of the procedure, a purely dissipative evolution acted for some time,
corresponding to replacement $i\longrightarrow -1$ in the left-hand side of Eq.(\ref{GP}).
At that, a fictitious short-ranged pinning potential was added to the harmonic trap potential
along the vortex line, and it prevented the vortex to move away from the prescribed curve.
Due to the dissipative stage, appreciable reduction of hard potential modes of the system 
on its main scale was achieved, which modes were inevitably excited after using
the mentioned rather arbitrary analytical formula. After that, the time was nullified, 
and the conservative dynamics in accordance with Eq.(\ref{GP}) was switched on.

For visualization of the vortex shape in the course of its motion, with definite time intervals,
coordinates of all those grid nodes where $|\psi|^2<0.05$ inside the ellipsoid 
$3/2-(r^2+az^2)/2=0.3$ were recorded. Clearly, such points are located inside the vortex core.
Using them, the dynamics was analyzed, and figures like Fig.1 were drawn.

\begin{figure}
\begin{center}
(a)\epsfig{file=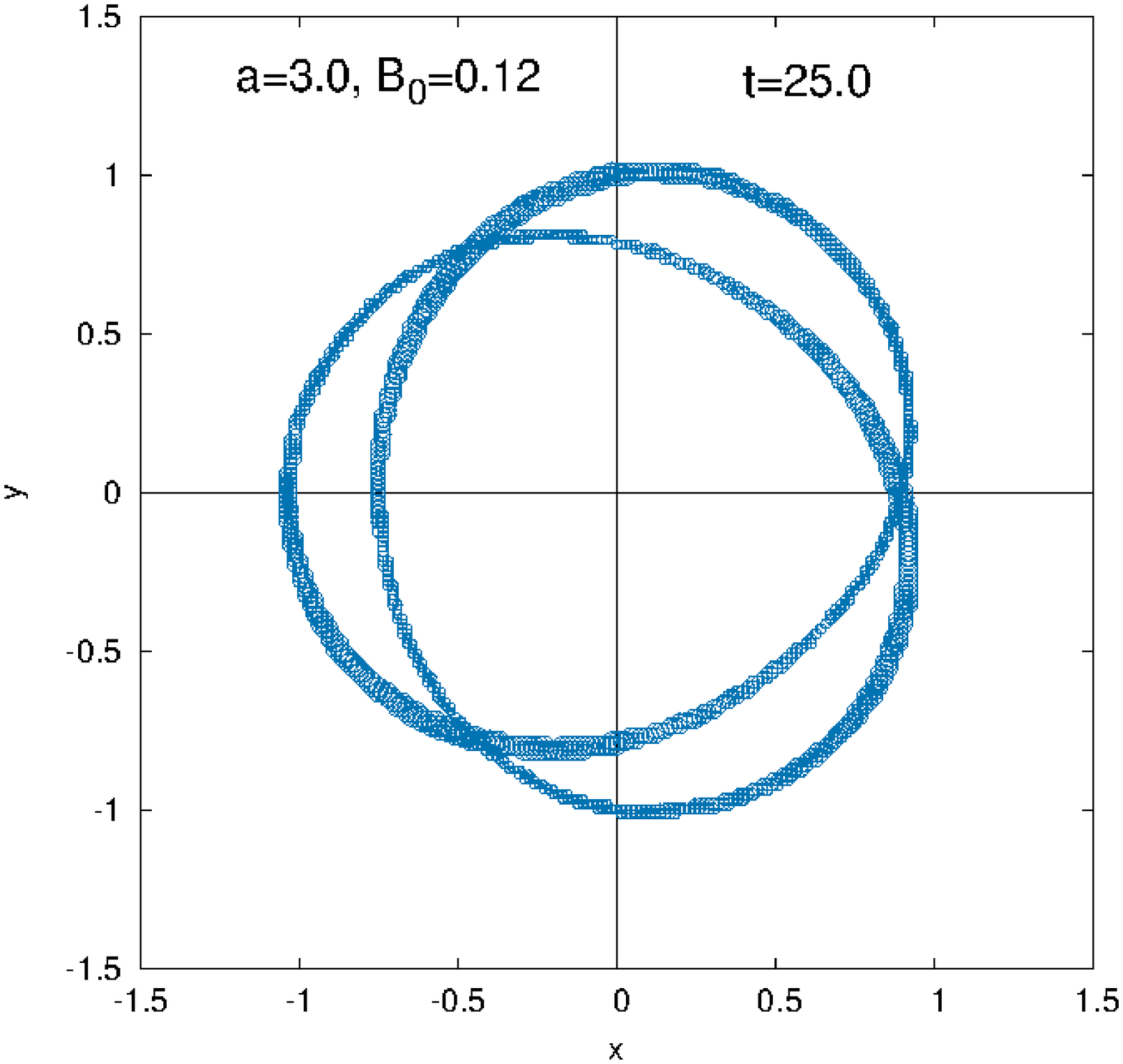, width=75mm}\\
(b)\epsfig{file=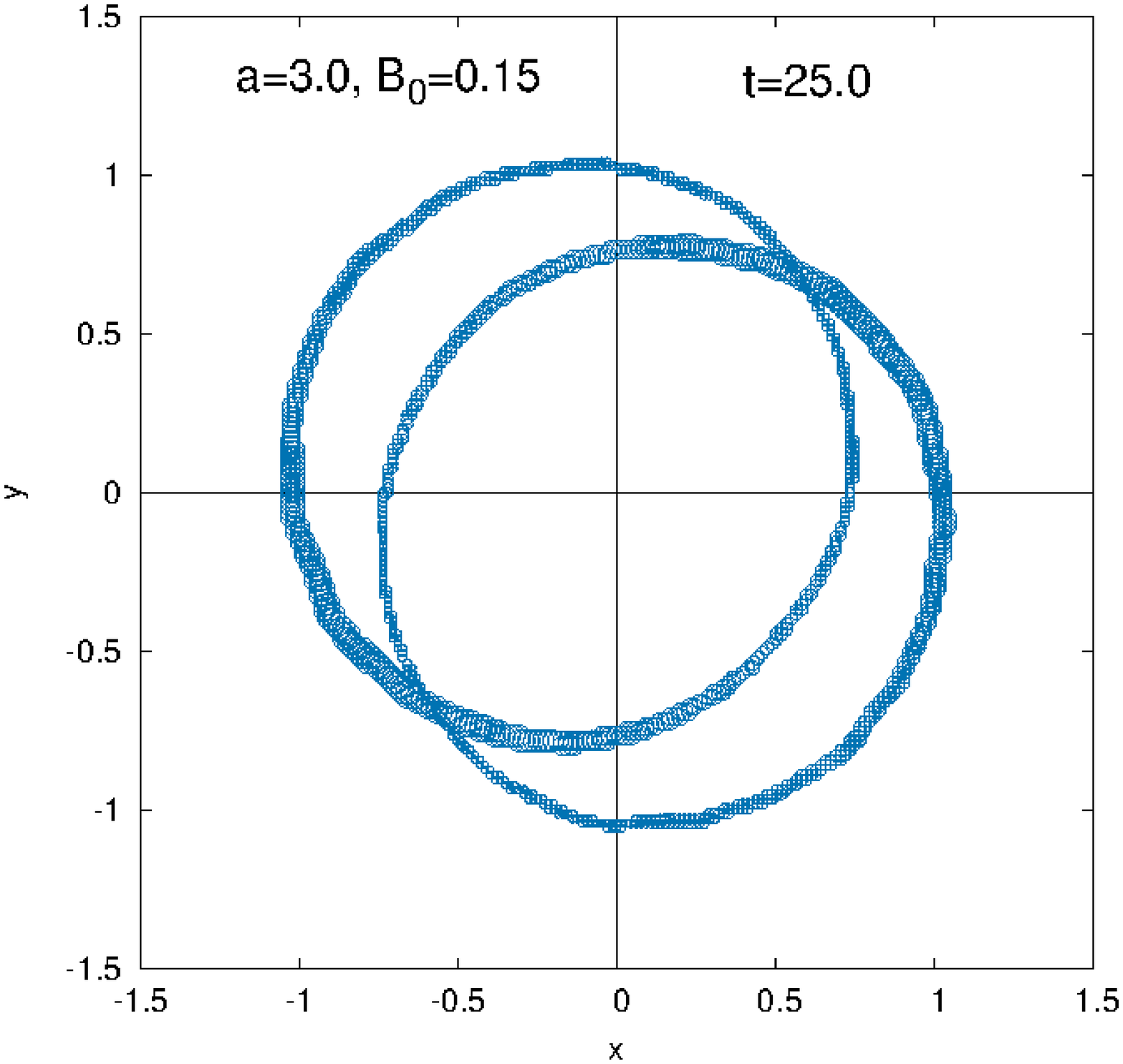, width=75mm}
\end{center}
\caption{Projections of quasi-stable vortex structures on $(x,y)$ plane in the case $a=3.0$,
at sufficiently large $t$: a) trefoil-knot; b) the simplest link of two rings.
The figures consist of  many small circles centered at the numerically found points 
inside vortex core. In order to determine the topology of vortex structure unambiguously, 
the radius of each such circle is made the larger, the larger value of the corresponding 
$z$-coordinate is.}
\label{Vortex_a20_B015} 
\end{figure}

\begin{figure}
\begin{center}
(a)\epsfig{file=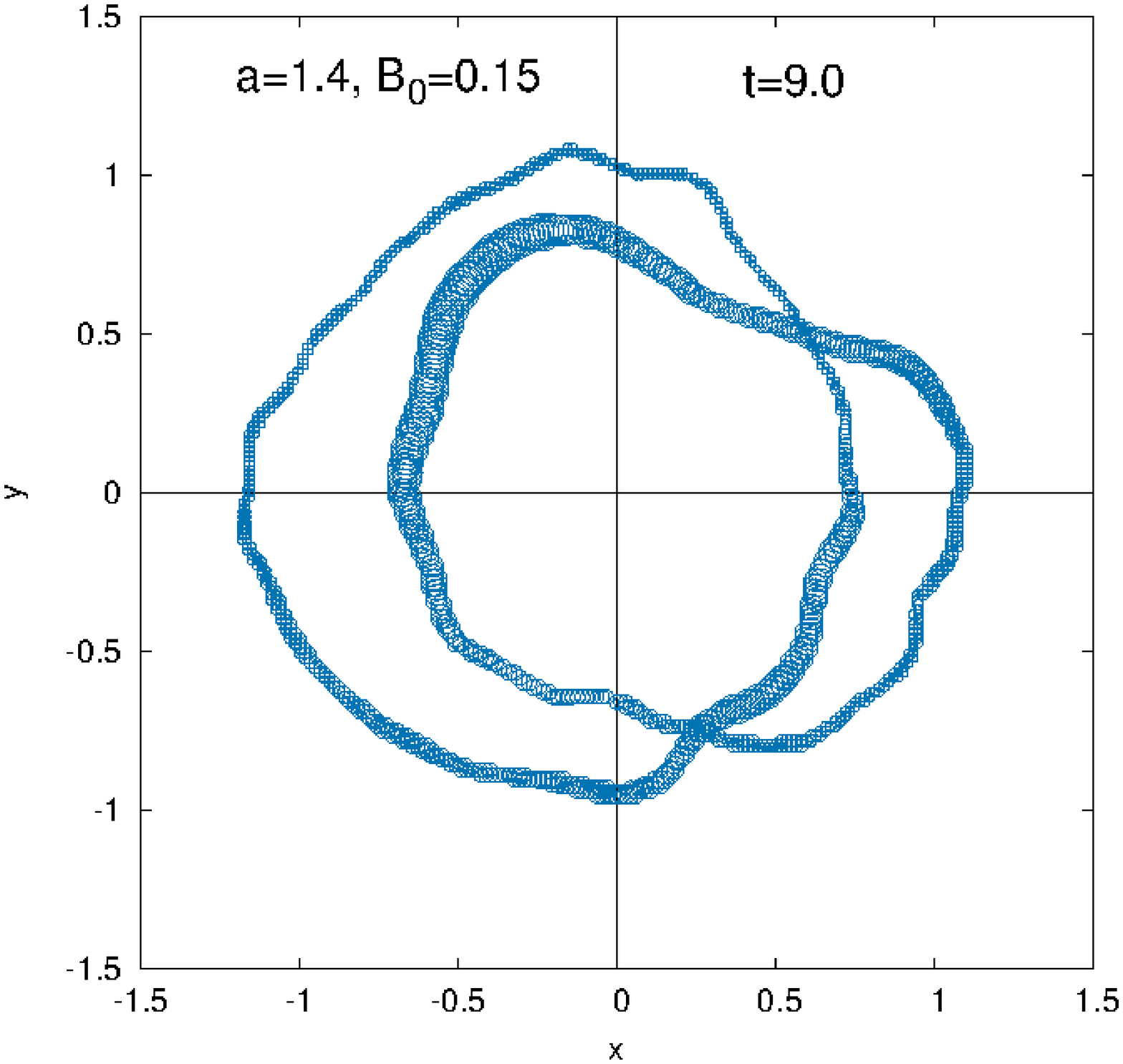, width=75mm}\\
(b)\epsfig{file=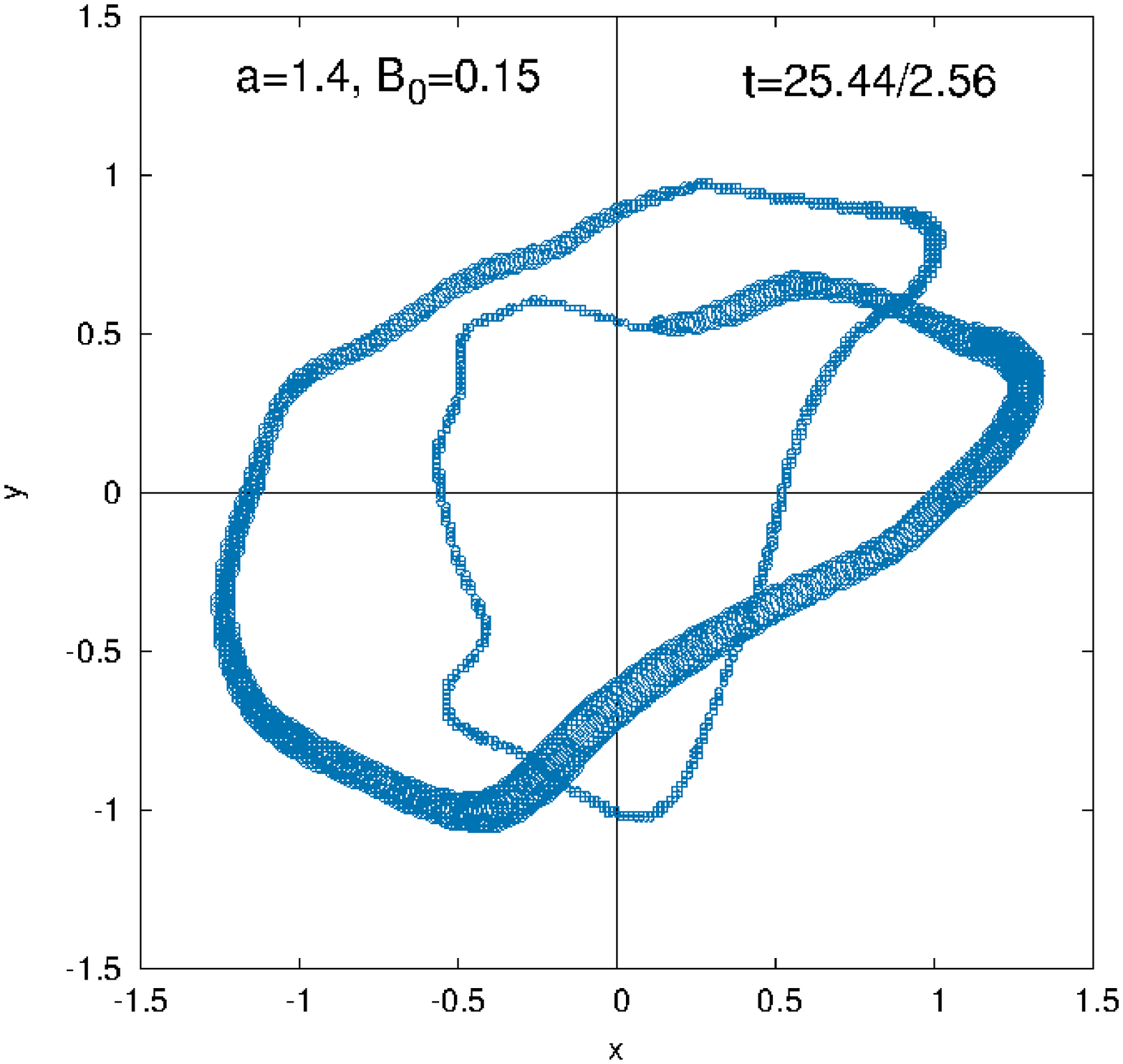, width=75mm}\\
(c)\epsfig{file=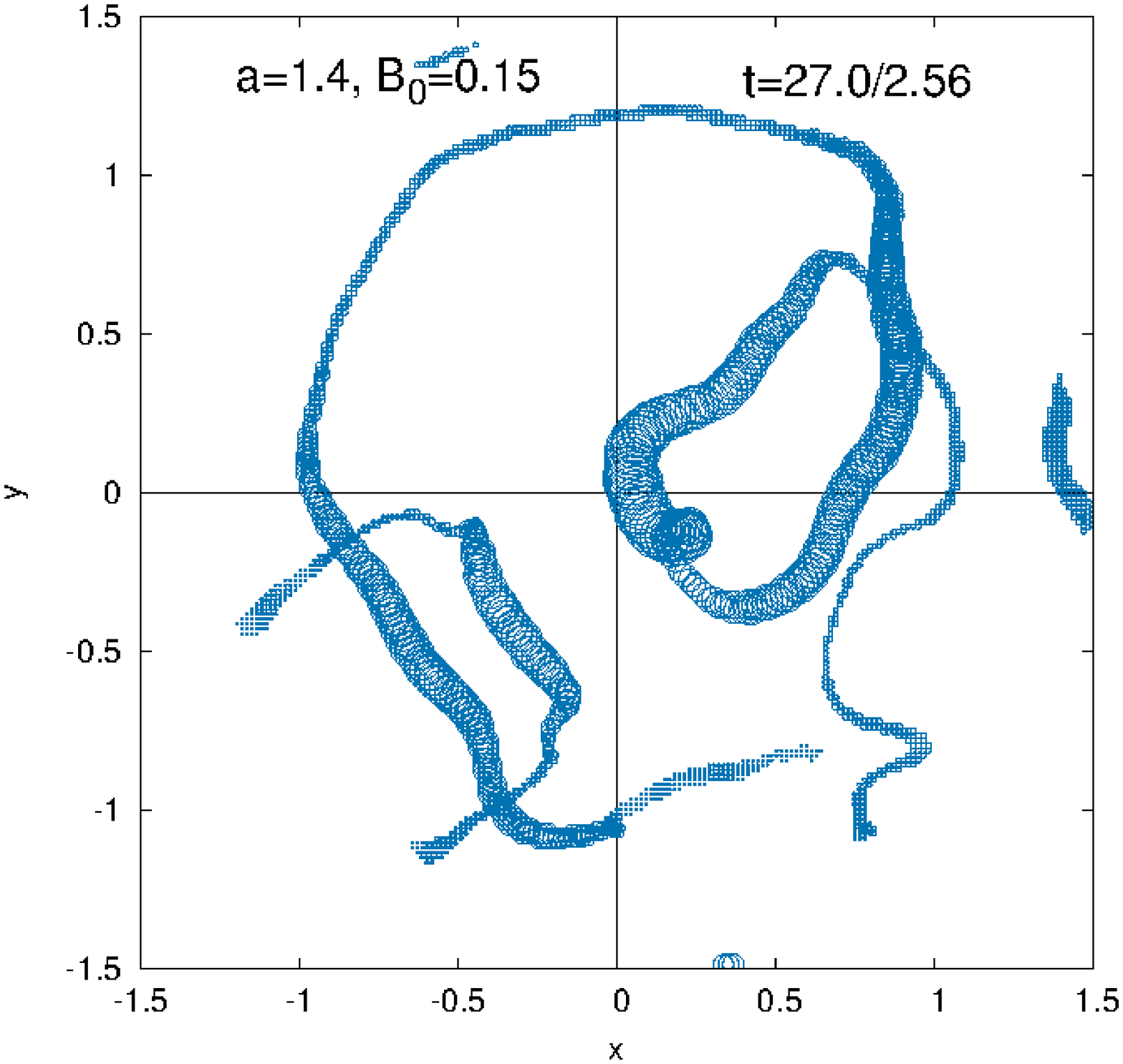, width=75mm}
\end{center}
\caption{The link of two rings turned out to be unstable in the case $a=1.4$:
a) strong deformation; b) change of the topology as the result of reconnection;
c) formation of vortex filaments with ends at the condensate surface.}
\label{Link_a14_B015} 
\end{figure}

{\bf Results}.
Numerical experiments were conducted for the pairs $(P,Q)=(2,3)$ --- (trefoil-knot),
$(2,2)$ --- the simplest link of two rings, and for $(2,1)$ --- (unknot in the form 
of a circle folded in two). The values $a$ were taken from the set $\{1.4, 2.0, 2.5, 3.0\}$,
while the parameter $B_0$ from the set $\{0.09, 0.12, 0.15, 0.18, 0.21\}$.
A quasi-stationary regime was observed very clearly for trefoil knot at $a=3.0$, $B_0=0.12$
and $a=2.5$, $B_0=0.18$, for the link of two rings at $a=2.5$, $B_0=0.18$, while
for the unknot at $a=2.5$, $B_0=0.21$. But in many other cases as well, deformation of
vortices was growing so slowly that a large life-time $T_{\rm life}\gtrsim 30$ was attained  
(for comparison, a typical vortex revolution time is about $\tau\approx 4\pi B_0^2\lesssim 0.3$)
It should be kept in mind that not all possible combinations have been tried yet, because 3D 
computations are not so fast, several days per one run. Of course, such investigation cannot be
considered as complete. A more detailed exploration of the parametric plane $(a,B_0)$ for
each kind of vortex structures will require massive computations with a fine step on parameter $B_0$.
At least, the experience in study of instabilities of vortex knots and links in a uniform 
superfluid \cite{R2018-2,R2018-3} confirms the necessity of small step $\Delta B_0\sim 0.001$.

In general, the numerical results have confirmed the possibility of existence of
long-lived quantum vortex knots and links within the Gross-Pitaevskii model with a 
quadratic external potential. In detail, however, some discrepancies between the approximate 
theory and the numerics were observed. Apart from the mentioned decrease of the mean radius
of quasi-stationary tori, actual parameters $\alpha$ and $\epsilon$ have changed too. 
As the result, even with  $a=3$, the instability sometimes developed (for example, 
at $B_0=0.15$ and $0.09$ in the trefoil case). On the other hand, at $a\neq 3$ the vortices 
were often quite long-lived, though in some runs excitaions in the form of Kelvin waves 
appeared on them after a time. And, of course, in every case potential motions (sound waves) 
took place in the condensate, looking as trembling ripples on the static density background.

Development of instability is illustrated in Fig.2 by the example of link of two rings.
First, deformation of unstable vortices is growing, and then they reconnect.
Since the main purpose of this work was a search for quasi-stationary structures, 
and in this sense the instability development is a negative result, we do not describe 
here the reconnection process in all detail. Let us only say that very quickly 
(but sometimes with a considerable delay) the subsequent evolution leads to approaching 
of some parts of vortices to the surface, their rupture and formation of vortex filaments.
The scenario is also possible when the rupture of the vortex at the surface occurs before 
its internal reconnection. After that the filaments decrease their lengths due to interaction 
with potential degrees of freedom, degenerate into half-rings, and ultimately leave the condensate.
In view of such tendency of vortices to exit from the condensate cloud, the found long-lived
quasi-stationary structures seem even more interesting objects.

{\bf Conclusions}.
The numerical experiments carried out in this work confirm clearly the recent hypothesis
about theoretical possibility of long existence of quasi-stationary, topologically 
nontrivial vortex structures in trapped Bose-Einstein condensates. A more detailed ascertainment 
of boundaries of quasi-stable domains will require massive computations. Let us also note 
that at $P\geq 3$ possible knots are not exhausted by torus knots. So, an interesting problem
for future research is to find long-lived configurations among such knots as the Figure-Eight Knot, 
the Granny Knot etc.

\end{document}